\documentclass[aps,prl,preprintnumbers,showpacs,twocolumn,nofootinbib]{revtex4} 
\usepackage{graphicx}
\usepackage{dcolumn}
\usepackage{bm}
\begin{document}
\newcommand{\Od}{{\cal O}}

\input epsf \renewcommand{\topfraction}{0.8}

\title{Confronting quintessence models with recent high-redshift 
supernovae data}
\author{G. Barro Calvo}
\affiliation{Departamento de Astrof\'{\i}sica, 
Universidad Complutense de
  Madrid, 28040 Madrid, Spain}
\author{A.~L.~Maroto}
\affiliation{Departamento de  F\'{\i}sica Te\'orica,
 Universidad Complutense de
  Madrid, 28040 Madrid, Spain}

\date{\today}

\begin{abstract}
We confront the predictions of different   quintessence models with
recent measurements of the luminosity distance 
from two sets of supernovae type Ia. In particular, we consider 
the 157 SNe Ia in
the Gold dataset  with $z<1.7$, and the more recent 
data containing 71 supernovae obtained by the Supernova Legacy Survey (SNLS)
with $z<1$. We numerically solve the evolution equations for
the Ratra-Peebles inverse power law model, the double exponential 
and the hyperbolic cosine quintessence models. We obtain confidence regions
from the two datasets in the $\Omega_M-\alpha$ and $\Omega_M-w_\phi$
planes for the different models and compare their predictions
with dark energy models with constant equation of state.    
\end{abstract}


\maketitle

\section{Introduction}
The
growing observational evidence from high-redshift supernovae 
\cite{Gold,SNLS}
and other cosmological data \cite{WMAP} suggests that the dominant
component of the universe today is some sort of dark energy 
fluid with negative pressure \cite{review,review2}. 
Indeed a component with constant equation of state $p_X=w_X\rho_X$ and 
 $w_X<-0.78$ \cite{WMAP}
fits the existing  data with reasonable goodness 
(XCDM model). 
In particular, it includes the most economical explanation for 
the present state of accelerated expansion of the universe, 
i.e., the presence of a pure cosmological
constant with $w_X=-1$. 
However, ther are other models in which the equation
of state depends on redshift and which are also able to fit the data with
comparable quality. This is the case of the so called quintessence
models \cite{RP,quintessence}, 
in which it is the presence of an evolving scalar field with
appropriate potential term what plays the role of dark energy.

Quintessence models exhibit an interesting property usually called 
 {\it tracking} behaviour \cite{tracking}. This means that  there is a 
wide range of initial conditions for which the evolution
of the scalar fields converges to a common evolutionary track.
Furthermore in such a tracking regime the equation of state
 of quintessence $w_\phi$ remains almost constant. Moreover there
are particular models, usually called {\it scaling} models \cite{scaling,scaling2,scaling3},  
in which $w_\phi$ mimics the equation of state of the dominant 
component throughout the cosmological evolution. These properties allow
quintessence models to alleviate the fine tuning problem of XCDM models, 
although in any case the scale of the
quintessence potential has always to be tuned in order to reproduce
the data. 

Apart from quintessence, other models have been proposed
in the literature. Thus for instance,  scalar fields with 
non-canonical kinetic terms (k-essence) \cite{k-essence};  
the generalized Chaplygin gas model \cite{Chaplygin}, 
which in principle allows for an unification of dark energy and dark 
matter;  infrared modifications of General Relativity, 
as for instance
in extra dimensional theories \cite{infrared} or the modification of the 
Friedmann equation in the so called Cardassian models \cite{Cardassian,Bento}. 

The present SNe Ia data are not very constraining 
when trying to determine the
nature of dark energy and different  dataset favor different regions
of the parameter space. 
Thus for instance, it is known that the Gold 
dataset \cite{Gold}
prefers models with $w_X<-1$ \cite{ness1,pref_phantom}, the so called phantom dark energy
models, whereas for the more recent 
Supernova Legacy Survey (SNLS) \cite{SNLS}, the best fit 
is closer to a pure cosmological constant \cite{Nesseris,pref_LCDM}.  
Therefore it is worth 
exploring what kind of constraints are imposed in each case, and this is
the aim of the present work. 
However the information we will obtain will be very
limited because even the highest redshift points in those sets 
explore only  relatively recent
epochs, and therefore the present supernovae data alone are not 
able to discriminate between 
the different models (although larger future 
supernovae catalogues are expected to improve in an important way 
the present constraints). Accordingly  we should  rely on
complementary information in order to distinguish  quintessence models
from a cosmological constant or other alternatives \cite{alter1,alter}. 
Thus for instance, in the  particular case of quintessence, 
dark energy is generated 
by a scalar field,
and there exists the  possibility of having density or velocity 
perturbations
which could affect  CMB anisotropies \cite{DEperturbations}. 
In constrast,  
the pure cosmological constant case does not support 
such perturbations.  

In this  work we will concentrate on different quintessence models 
and derive constraints
on their parameters from the two mentioned Gold and SNLS  
datasets. We consider three models:  an inverse power-law 
Ratra-Peebles (RP) model \cite{RP}, 
$V(\phi)=M^{4+\alpha}/\phi^\alpha$, the double
exponential (DE) potential $V(\phi)=M^4(e^{\alpha \phi}
+e^{\beta\phi})$ \cite{DE,DE2}
which exhibits scaling behaviour and the hyperbolic cosine (HC) $V(\phi)=
M^4(\cosh(\lambda\phi)-1)^\alpha$ \cite{HC}, 
which posseses an oscillatory behaviour 
at recent epochs. Unlike previous works \cite{param_w, ichikawa}, we do not parametrize the
equation of state $w_\phi(z)$ in order to obtain explicit expressions
for the luminosity distance-redshift relation $d_L(z)$. Instead, we solve
numerically the evolution equation for the scalar field and the
universe scale factor. This allows us to numerically obtain
 $d_L(z)$ for each model. The value of the fitted cosmological 
parameters is known to have a strong dependence on the 
particular parametrization chosen for the equation of
state (see \cite{bassett,review2} and references therein). 
In particular, at least three parameters are needed in
order to take into account properly all the information
encoded in the SNIa data. Our approach does not rely 
on particular parametrizations, but instead we consider the
full redshift dependence by numerically solving the 
evolution equations. 

The plan of the paper goes as follows, in section 2 we review 
the models properties
and obtain the appropriate equations of motion. In section 3, we compute
the luminosity distance expressions and perform the corresponding 
cosmological fits. Section 4 contains the main results of the paper
with the confidence regions for different spaces of parameters and finally
we include a brief section with conclusions.

\section{Quintessence models and evolution equations} 

\subsection{Inverse power law potential}
In the first model that we will consider, the potential reads:
\begin{eqnarray}
V(\phi)=\frac{M^{4+\alpha}}{\phi^\alpha}
\end{eqnarray} 
This model has a tracker attractor, but it is not scaling, although the
equation of state is determined by that of the dominant component:
\begin{eqnarray}
w_\phi=\frac{\alpha w_{R,M}-2}{\alpha+2}
\end{eqnarray} 
where, for $\alpha>-2$,  
in the radiation or matter eras we have $w_\phi<w_{R,M}$ 
respectively, i.e. the energy density decreases more slowly than the
dominant component and it eventually becomes dominant at late times.
The corresponding equations of motion reduce to the Friedmann equation for
the scale factor, together with the scalar field evolution equation:
\begin{eqnarray}
H^2&=&\frac{8\pi G}{3}\left(\frac{\dot\phi^2}{2}+V(\phi)+\sum_{i=M,R} 
\rho_{i,0}a^{-3(w_i+1)}\right)\nonumber \\
\ddot\phi&+&3H\dot\phi+V'(\phi)=0
\end{eqnarray}
where $\rho_{i,0}/\rho_0=\Omega_i$ with $i=M,R$ and $\rho_0$ is the
critical density. Here a dot denotes derivative with respect to 
cosmological time.

In order to solve these equations numerically, we transform them into
dimensionless equations, defining: $\tau=H_0t$ and 
$\tilde \phi=(8\pi G)^{1/2}\phi$. Now a prime will denote derivative 
with respect to $\tau$:
\begin{eqnarray}
\frac{\tilde\phi''}{3}&+&\tilde\phi'\frac{a'}{a}+\tilde V'(\tilde\phi)=0
\nonumber \\
\left(\frac{a'}{a}\right)^2&=&\frac{\tilde\phi'^2}{6}+\tilde V(\tilde\phi)+
\Omega_R a^{-4}+\Omega_M a^{-3}\label{eqs}
\end{eqnarray} 
where
\begin{eqnarray}
\tilde V(\tilde \phi)&=&\frac{A}{\tilde \phi^\alpha}\nonumber\\
A&=&\frac{M^{4+\alpha}(\sqrt{8\pi G})^{\alpha +2}}{3H_0^2}
\end{eqnarray}
The initial conditions for $\tilde\phi$ and $a$  will be given 
by: $\tilde\phi(0)=5\cdot 10^{-1}$, $\tilde\phi'(0)=0$
and $a(0)=10^{-3}$. We will explore the parameters
range $\alpha=0-5.5$ and $A=0.3-8\cdot 10^4$. Notice that 
for those values of the parameters, the initial 
scalar field energy density  is a small fraction of $\rho_{M}$ and
$\rho_R$. Notice also that 
thanks to the attractor behaviour, the late time evolution
is not very sensitive to the particular initial conditions chosen.

\subsection{Double exponential potential}

In this case:
\begin{eqnarray}
V(\phi)=M^4(e^{\alpha\kappa\phi}+e^{\beta\kappa\phi})
\end{eqnarray} 
where $\kappa=\sqrt{8\pi G}$. The energy density in this model 
follows a scaling behaviour at early time, whereas it becomes dominant
at late times with appropriate equation of state: $w_\phi\simeq -1$.
The corresponding dimensionless potential now reads:
\begin{eqnarray}
\tilde V(\tilde \phi)&=&A(e^{\alpha\tilde\phi}+e^{\beta\tilde\phi})
\nonumber\\
A&=&\frac{M^4(8\pi G)}{3H_0^2}
\end{eqnarray}
and the corresponding equations of motion are given in (\ref{eqs}).
The initial conditions  are $\tilde\phi(0)=0.5$, $\tilde\phi'(0)=0$
and $a(0)=10^{-3}$. The range of parameters considered are $\beta=20$,
$\alpha=0.1-1.5$ and $A=0.4-120$, where
for simplicity we have fixed the constant $\beta$. 
The $\alpha$ and $\beta$ constants essentially
fix the value of $w_\phi$ today, and we have checked that fixing 
$\beta$ still allows us to cover a wide range of values in $w_\phi$, 
just varying $\alpha$.
 Again for those particular values,  
the initial energy density in the scalar field is a small fraction
of the radiation and matter densities. 

\subsection{Hyperbolic cosine potential}

In this case the dimensionless potential reads:
\begin{eqnarray}
\tilde V(\tilde \phi)&=&A(\cosh(\lambda\tilde\phi)-1)^\alpha
\nonumber\\
A&=&\frac{M^4(8\pi G)}{3H_0^2}
\end{eqnarray}
In the limit $\lambda\tilde\phi\gg 1$, i.e. at early times,  
the potential behaves as
a single exponential and it is possible to choose the parameters
so that the model posseses a scaling behaviour. In the opposite limit, 
$\lambda\tilde\phi\ll 1$ (late times), the potential can be approximated by
$\tilde V(\tilde \phi) \propto (\lambda\tilde\phi)^{2\alpha}$, and the 
scalar field oscillates around  $\tilde\phi=0$. When oscillations start, 
the equation of state no longer mimics the dominant component, but it
also start oscillating with average value given by:
\begin{eqnarray} 
\langle w_\phi \rangle=\frac{\alpha-1}{\alpha+1}
\end{eqnarray}
Accordingly, the value of $\alpha$ determines the present equation
of state.  For $\alpha=1$ the oscillations behaves as nonrelativistic 
matter, whereas if we require accelerated expansion today, i.e. 
$w_\phi<-1/3$ then we should have $\alpha<1/2$. 

The numerical integration of equations (\ref{eqs}) now has an additional
difficulty. In the interesting case ($\alpha<1/2$), 
the potential derivative $\tilde V'$
appearing in the first equation diverges at the origin. Since both 
$\tilde\phi$ and $\tilde\phi'$ are continuous at the origin, in order
to avoid numerical  instabilities we have smoothed the divergence
modifiying the potential as 
$V(\tilde \phi)=A(\cosh(\lambda\tilde\phi)-(1-\epsilon))^\alpha$.
We have checked that for sufficiently small values, the 
results do not depend on the $\epsilon$ parameter.

The initial conditions considered are $\tilde\phi(0)=1.6$, 
$\tilde\phi'(0)=0$, $a(0)=10^{-3}$. The free parameters are 
$\alpha=0.02-0.4$, $A=0.3-14$ and we have fixed $\lambda\alpha=5$, 
which ensures again that initially quintessence 
is not the dominant component of the energy density and that it
follows a scaling behaviour.

\section{Cosmological data fit}
After numerically solving  (\ref{eqs}), 
we obtain a discrete time evolution for the scale
factor $a(t)$ and the scalar field $\phi(t)$ for a given 
set of parameters $(A,\alpha)$.  
The time dependence will be used to compute the theoretical 
values of the luminosity distance to each SNe Ia 
with a given redshift \emph{z}. Thus we use
 the well known expression, derived 
from the FRW metric under flat prior, 
expressed in terms of the dimensionless time variable $\tau$.
\begin{equation} \label{lum_dist}
d_{L}(\tau)=\frac{a(\tau_{0})}{a(\tau_{1})}\int_{\tau_{0}}^{\tau_{1}}
\frac{d\tau}{a(\tau)}
\end{equation}

The observations of supernovae measure essentially the distance 
modulus $\mu$, which is the diference between the 
apparent magnitude \textit{m} and the absolute magnitude \textit{M}, 
and relates to the luminosity distance as:
\begin{eqnarray}
\mu_{th}=m-M&=&5\log{d_{L}}+5\log\left(\frac{cH_{0}^{-1}}{\mbox{Mpc}}\right)
+25
\nonumber  \\
&=&5\log{d_{L}}+\tilde{M}
\end{eqnarray}
The $M$ value can be assumed constant once the necessary 
corrections are applied on $m(z)$.

Assuming $\tau_{0}$ to be the value for which $a(\tau_{0})=1$, the 
scale factor depends on redshift as $a(\tau)=(z+1)^{-1}$. 
As a consequence, for each supernovae in a given set, we derive 
$a(\tau_{1})$ from its redshift and obtain $\tau_{1}$ from the 
numerical output data of (\ref{eqs}). Hence the integral 
(\ref{lum_dist}) can be numerically evaluated to obtain $\mu_{th}$ 
for each ($A$, $\alpha$), once the value of $H_{0}$ is fixed.

Once we obtain $\mu_{th}$, the comparison to its observational
 value will enable us to carry out a $\chi^2$ statistical analysis.
 For this purpose, we have considered two sets of supernovae. 
On one hand, the Gold set, compiled by Riess et. al. \cite{Gold}, 
containing 143 points from previously published data, 
plus 14 points with  $z>1$  discovered with the HST, 
all reduced under the same criteria in order to improve the errors 
arising from systematics. On the other, 
the SNLS set, comprising 71 distant 
supernovae ($0.15 < z < 1$) discovered during the first year of the 
Supernova Legacy Project (SNLS) \cite{SNLS}, alongside with 44 SNe Ia 
from other sources that feeds the nearby zone ($0.0015 < z < 0.125$), 
and which are also included in the Gold set.

The process followed to obtain $\chi^{2}$ is slightly 
different for each set. In the Gold set we 
have used the observational distance modulus 
$\mu_{exp}$ given by Riess et. al. together with 
its associated error $\sigma_{\mu}$ to compute $\chi^{2}$ as a 
function of the free parameters of the quintessence model.

\begin{equation}
\chi^{2}(A,\alpha,\tilde{M})=\sum_{i=1}^{N}
\frac{(\mu_{obs}-\mu_{th}(A,\alpha,\tilde{M}))^{2}}{\sigma_{i\mu}^{2}}
\label{goldchi}
\end{equation}
\noindent
The dependence on $H_{0}$ has been accounted for in (\ref{goldchi})
through the nuissance parameter $\tilde{M}$,
which is independent of the data points and the data set.
We have marginalized $\chi^2$ over all values of $\tilde{M}$
by expanding and minimizing (\ref{goldchi}) with respect to $\tilde{M}$
(see \cite{Bento,Nesseris,ness1}).

On the other hand, for the SNLS data, a more detailed relation 
between $\mu$ and the observational measurements has to be considered 
in the statistical analysis, which implies recursively 
fitting two new non-cosmological parameters in 
the calculation of $\chi^{2}$.

The expression for the observational distance modulus used 
by the SNLS team includes these two parameters to measure the 
impact of the rest frame color parameter ($c$), 
and the light curve strecht ($s$), on the distance modulus.

\begin{equation}
\mu^{SNLS}_{obs}=m(z)-M+a_{1}(s-1)-a_{2} c
\end{equation}
Introducing this equation into the expression for $\chi^{2}$ 
\begin{eqnarray}
&\left.\right.&\chi^{2}(\alpha,A,a_{1},a_{2},M+\tilde{M})= \\
&\sum_{i=1}^{N}&
\frac{(\mu_{obs}^{SNLS}(a_{1},a_{2},M)-\mu_{th}
(A,\alpha,H_{0},\tilde{M}))^{2}}{\sigma_{i\mu}^{2}}\nonumber
\end{eqnarray}
we obtain a six parameter dependence 
(five parameters in practice, since we can replace $M+\tilde{M}$ 
by a single additive constant) 
We minimize following the process suggested in \cite{SNLS}, i.e. 
marginalizing with respect to $a_1$, $a_2$ and $M+\tilde{M}$ 
for each pair of $(A,\alpha)$ values. Notice that marginalizing
over $M$ or $M+\tilde{M}$ is equivalent to marginalize with respect
to $H_0$. 
Again $\mu$ and $\sigma_{\mu}$ are obtained from the data in \cite{SNLS}.

\section{Results and Model comparison}

The best resulting cosmologies obtained after 
fitting each model to the Gold and SNLS sets are summarized  
in Table I and II respectively. 
In each sample, we find the same $\chi^{2}$ 
for the best fit with the three quintessence models considered in the 
paper, i.e.  $\chi^{2}_{min}=177.0$ for the 157 SNe Ia of the 
Gold set, and $\chi^{2}_{min}=111.0$ for the 
115 SNe Ia of the SNLS set. 
In addition, the minimun $\chi^{2}$ for all models 
corresponds to $\alpha=0$. For this value
the RP and HC  potentials turn into a cosmological 
constant term and we have used this fact to check our results, 
finding good agreement for the $\Omega_{M}$ value 
when compared to the Riess et. al., 
and Astier et. al. XCDM model. 
On the other hand, albeit the DE potential 
does not strictly behave as a 
$\Lambda$ term, due to the nonvanishing exponential, 
the $\omega_{\phi}$ value tends asymptotically to -1, 
leading to the same result as DE and HC.
This implies that a quintessence potential, 
and a pure $\Lambda$CDM model, can not be distinguished 
only by fitting  $(\Omega_{M},\omega_{\phi})$, since 
a quintessence model can be tuned to accurately 
resemble a cosmological constant, 
at least in the redshift interval explored by the actual SNe Ia sets. 
In any case, notice that $\rho_{\phi}$  evolves in time
for any non-zero value of  $\alpha$ for the three potentials.

\begin{table}[h]
\centering
\begin{tabular}{|c|c|c|c|c|}
\cline{2-5}  
\multicolumn{1}{c}{} & \multicolumn{4}{|c|}{Gold} \\
\cline{2-5} 
 \multicolumn{1}{c|}{} & $\Omega_{M}$   &  $\omega_{\phi}$ &  $\alpha$ & $\chi^{2}$\\
\hline 
RP  & 0.30$\pm^{0.06}_{0.11}$  &  -1$\pm^{0.24}_{\;-}$ & 0$\pm^{1.20}_{\;-}$ & 177.0 \\
HC  & 0.30$\pm^{0.06}_{0.10}$  &  -1$\pm^{0.24}_{\;-}$ & 0$\pm^{1.35}_{\;-}$  & 177.0 \\
DE  & 0.30$\pm^{0.06}_{0.07}$  &  -1$\pm^{0.22}_{\;-}$ & 0$\pm^{1.08}_{\;-}$ & 177.0 \\
\hline
\end{tabular}
\caption{\label{tabla_valores_cosmologicosI} Best fitting 
cosmological parameters, with their associated 
$68\%$ errors for the Gold set. The degeneracy 
in $\alpha=0$ leads to the same value of 
$\Omega_{M}$ for  the three potentials in a set. 
Moreover, due to the form of the quintessence Lagrangian 
values of $\omega_{\phi}$ below -1 cannot be obtained.}
\end{table}

Combining the data obtained by solving (\ref{eqs}) 
for each potential, we may relate ($A,\alpha$), and 
its corresponding fitting $\chi^{2}$, 
to a single ($\omega_{\phi},\Omega_{M}$) 
pair which allows us to plot the confidence regions for the 
two parameter combinations $(\alpha,\Omega_{M})$ and 
$(\omega_{\phi},\Omega_{M})$.

For the HC model, we  
use directly $\langle \omega_{\phi} \rangle=\frac{1-p}{1+p}$, 
instead of the value derived from eq.(\ref{eqs}), 
due to the difficulties arising from the calculation of 
$\langle \omega_{\phi} \rangle^{num}$ in the cases when 
the oscillations fails to complete a whole period. 
Moreover, a comparison between 
$\langle \omega_{\phi} \rangle$ and 
$\langle \omega_{\phi} \rangle^{num}$ shows  a small average discrepancy, 
$\Delta\omega_{\phi}=10^{-3}$, which justifies our approximation, 
and serves as a check  of the numerical solution.

\begin{table}[h]
\centering
\begin{tabular}{|c|c|c|c|c|}
\cline{2-5}  
\multicolumn{1}{c}{} & \multicolumn{4}{|c|}{SNLS} \\
\cline{2-5} 
 \multicolumn{1}{c|}{} & $\Omega_{M}$   &  $\omega_{\phi}$ &  $\alpha$ & $\chi^{2}$\\
\hline 
RP  &  $<$0.36  &   $<$-0.58 &  $>$0    &111.0 \\
HC  &  $<$0.36  &   $<$-0.61 &  $<$0.3  &111.0 \\
DE  &  $<$0.36  &   $<$-0.54 &  $<$1.27 &111.0 \\
\hline
\end{tabular}
\caption{\label{tabla_valores_cosmologicosII} Best fitting 
cosmological parameters, and upper bounds at $95\%$ C.L. for the SNLS set. 
The constraints on the cosmological paramaters hold for $\Omega_{M}>0.06$ 
for the RP and DE potentials, and $\Omega_{M}>0.12$ for the HC.}
\end{table}

Fig.1 to Fig.3 show the confidence regions for the models 
in the $\alpha-\Omega_{M}$ and $\omega_{\phi}-\Omega_{M}$ planes, 
together with the XCDM contour plots. 
The XCDM regions have been calculated using the same code on the 
Friedmann equation with a dark energy term ($\rho_{X},\omega_{X}$) 
instead of a coupled scalar field. In the $\alpha-\Omega_{M}$ 
plots the contours grow asymptotically for decreasing 
$\Omega_{M}$ \cite{caresia}.
Nevertheless, as $\omega_{\phi}$ follows the same 
growing trend, it is possible to restrict the maximum value of 
$\omega_{\phi}$. 

At the $95\%$ confidence level and for $\Omega_{M}>0.1$, 
we find $\omega^{gold}_{\phi}<$ -0.56, $\omega^{SNLS}_{\phi}<$ -0.58, 
for the RP model, and $\omega^{gold}_{\phi}<$ -0.55, 
$\omega^{SNLS}_{\phi}< -0.56$, for the DE model.
The HC potential also shows similar results for  
$\Omega_{M}>0.15$, limiting $\omega^{gold}_{\phi}<$ -0.58, and $\omega^{SNLS}_{\phi}<$ -0.64. Notice that we do not extract any result for values below $\Omega_M=$0.1 for RP and DE or
$\Omega_{M}=$0.15 for HC, 
since the computational effort increases in  those regions. In fact 
in the HC model that region is  also 
highly sensitive to the value of the smoothing factor, $\epsilon$.

By comparing the plots for the two datasets, 
we see that the width of the contours is essentially the same. 
However, the SNLS $68\%$ C.L. contour never closes for 
$\Omega_{M}>$0.06, whereas the Gold 
set constraints the cosmological parameters, 
for all models, 
within 0.25$<\Omega_{M}<$0.36 
and -1$<\omega_{\phi}<$-0.75, at this significance level. 
For the RP potential, these results can be compared to  analysis
done with previous supernovae data in \cite{alter1,caresia}.

Concerning the XCDM contours, one remarkable feature 
is that as $\omega_{\phi}\mapsto-1$,  
the SNLS confidence regions for quintessence 
and XCDM tends to overlap. This is an expected fact, since 
the best fitting value for XCDM according to Astier et. al., 
is $\omega_{\phi}=$-1.02, quite close to the quintessence 
best fit, which is precisely the degenerate case $\alpha=0$. 
On the other hand, the best fitting value for the 
Gold set, assuming no priors, 
is $\omega_{\phi}=$-2.3. This implies an important difference 
in $\chi^{2}$ with respect to the quintessence models, 
which prevents the confidence regions from matching. 
In spite of this, it is noticeable the similarity 
between contours given the fact that the $68\%$ C.L. region of XCDM 
is out of the plotted interval. Finally, 
for all models, the quintessence $99\%$ contours place 
an outer boundary containing all the XCDM regions, 
implying that any of the XCDM cosmologies, 
might be obtained from a quintessence potential.

\begin{figure*}[h]
\includegraphics[width=0.49\textwidth]{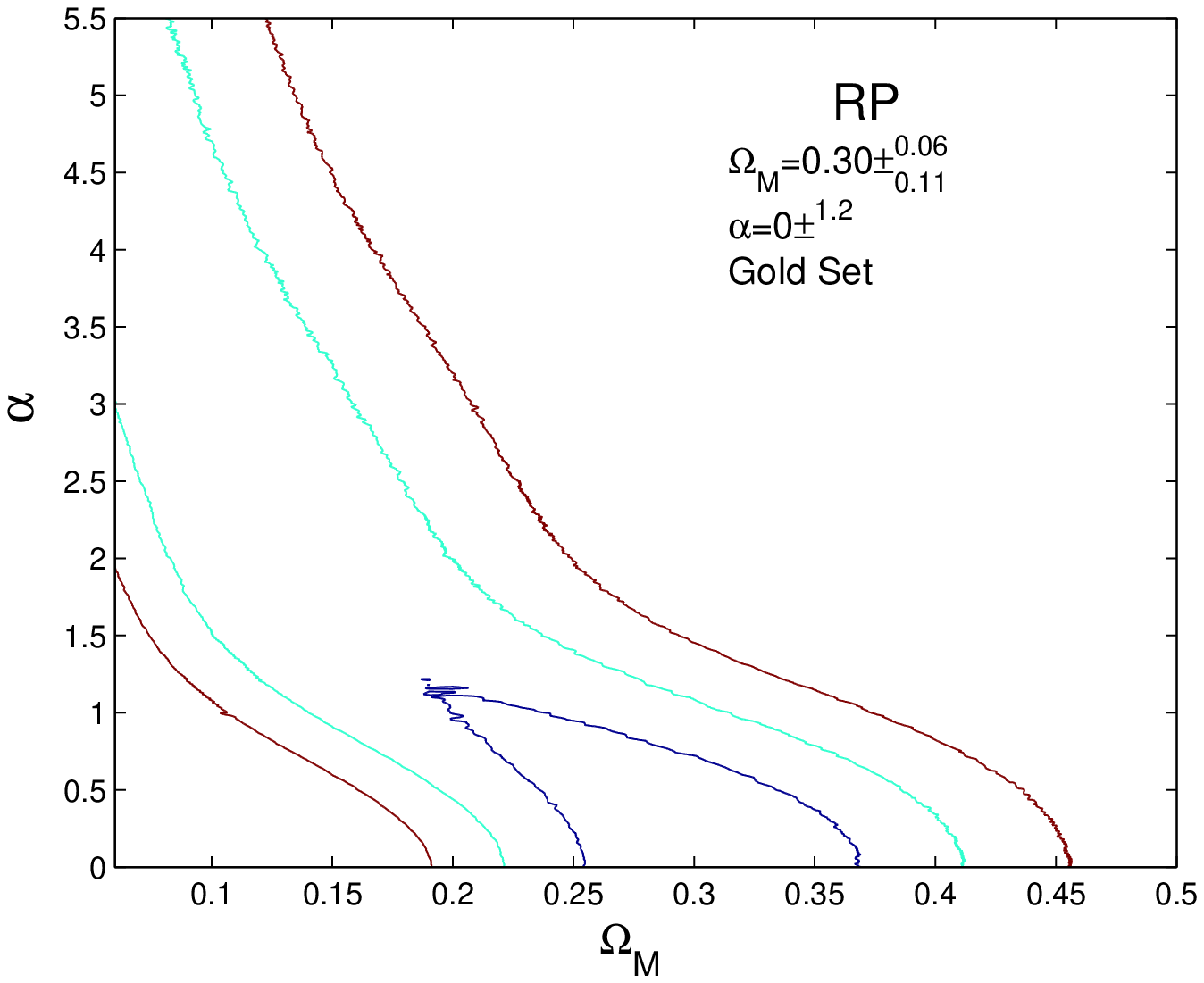}
\includegraphics[width=0.49\textwidth]{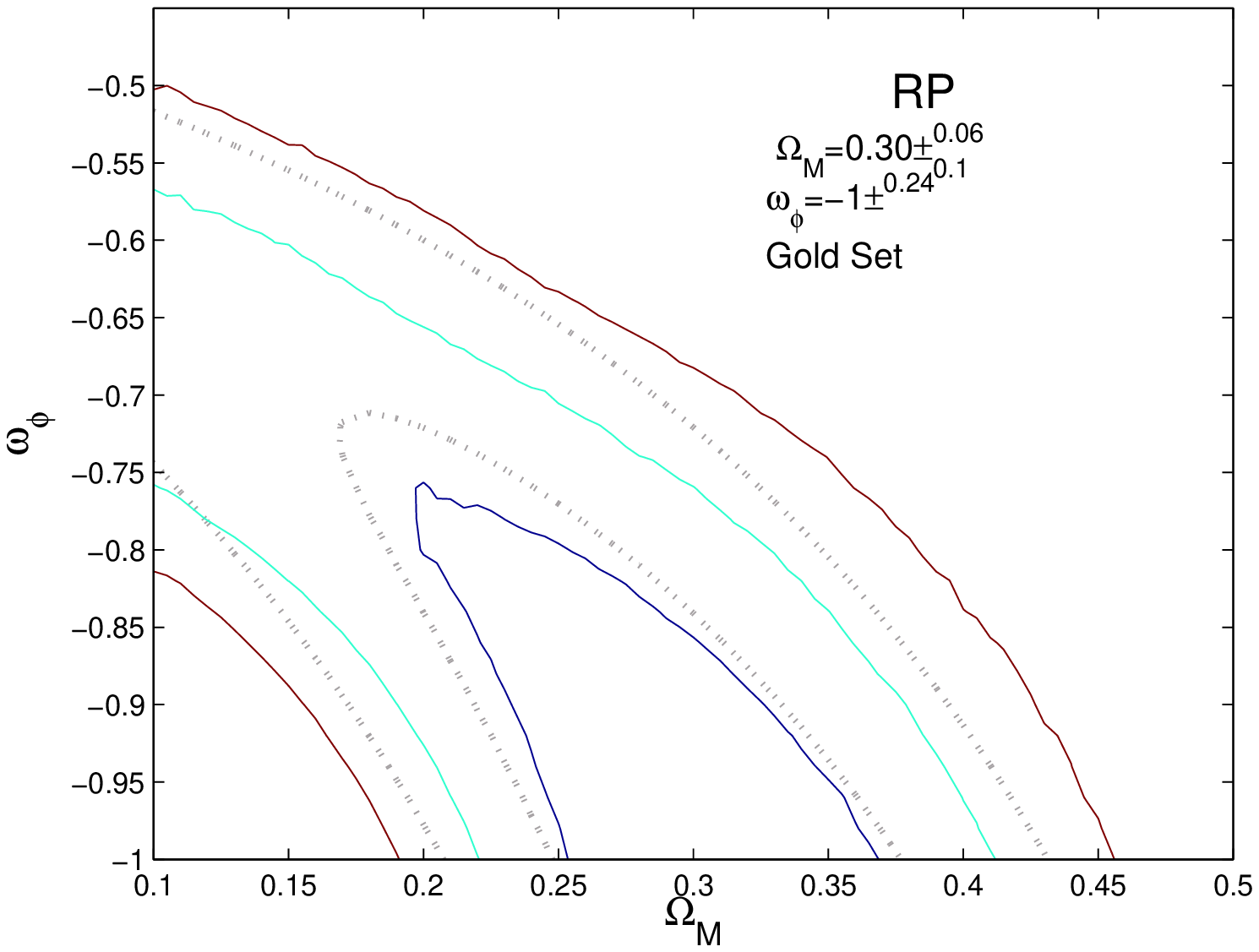}
\includegraphics[width=0.49\textwidth]{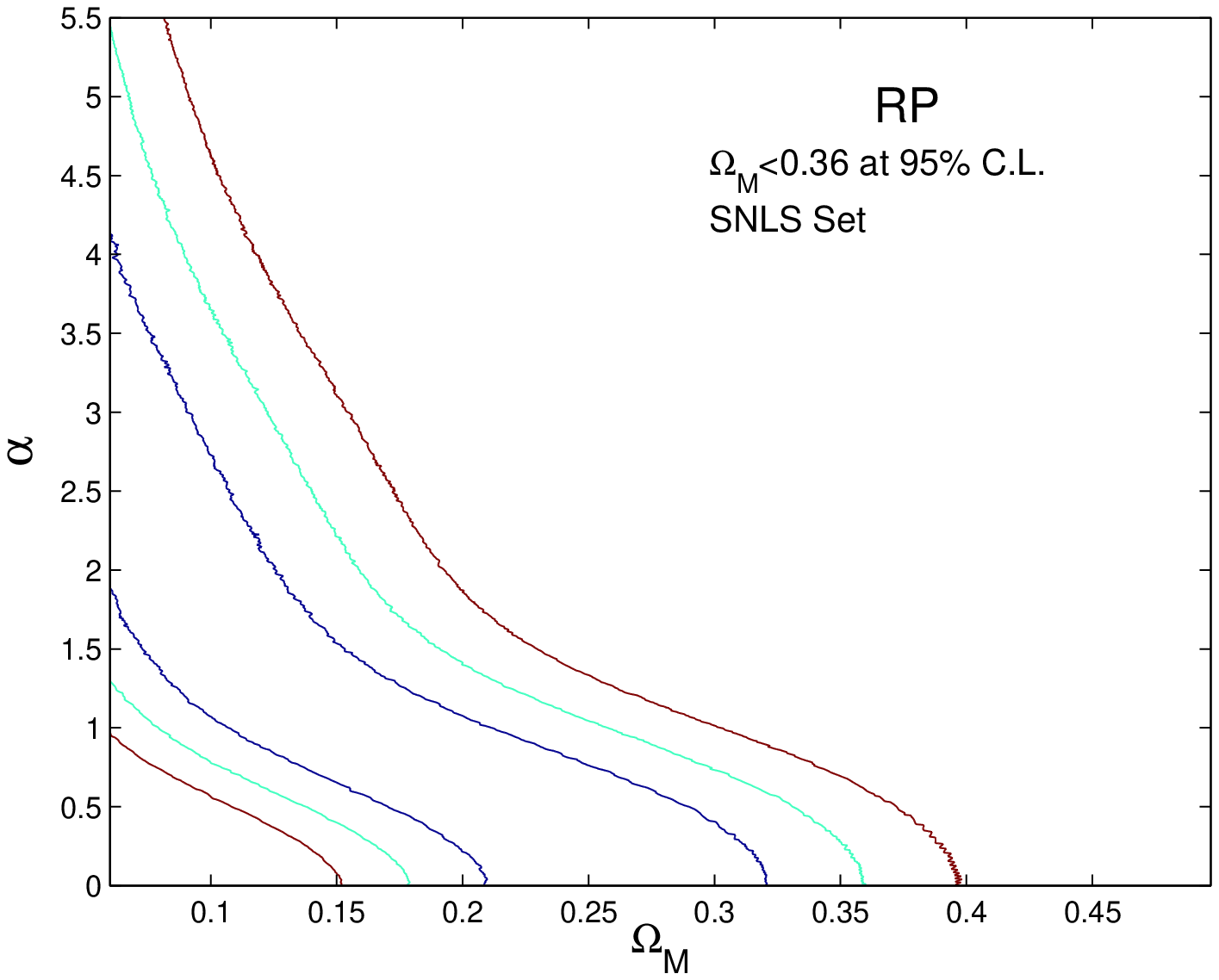}
\includegraphics[width=0.49\textwidth]{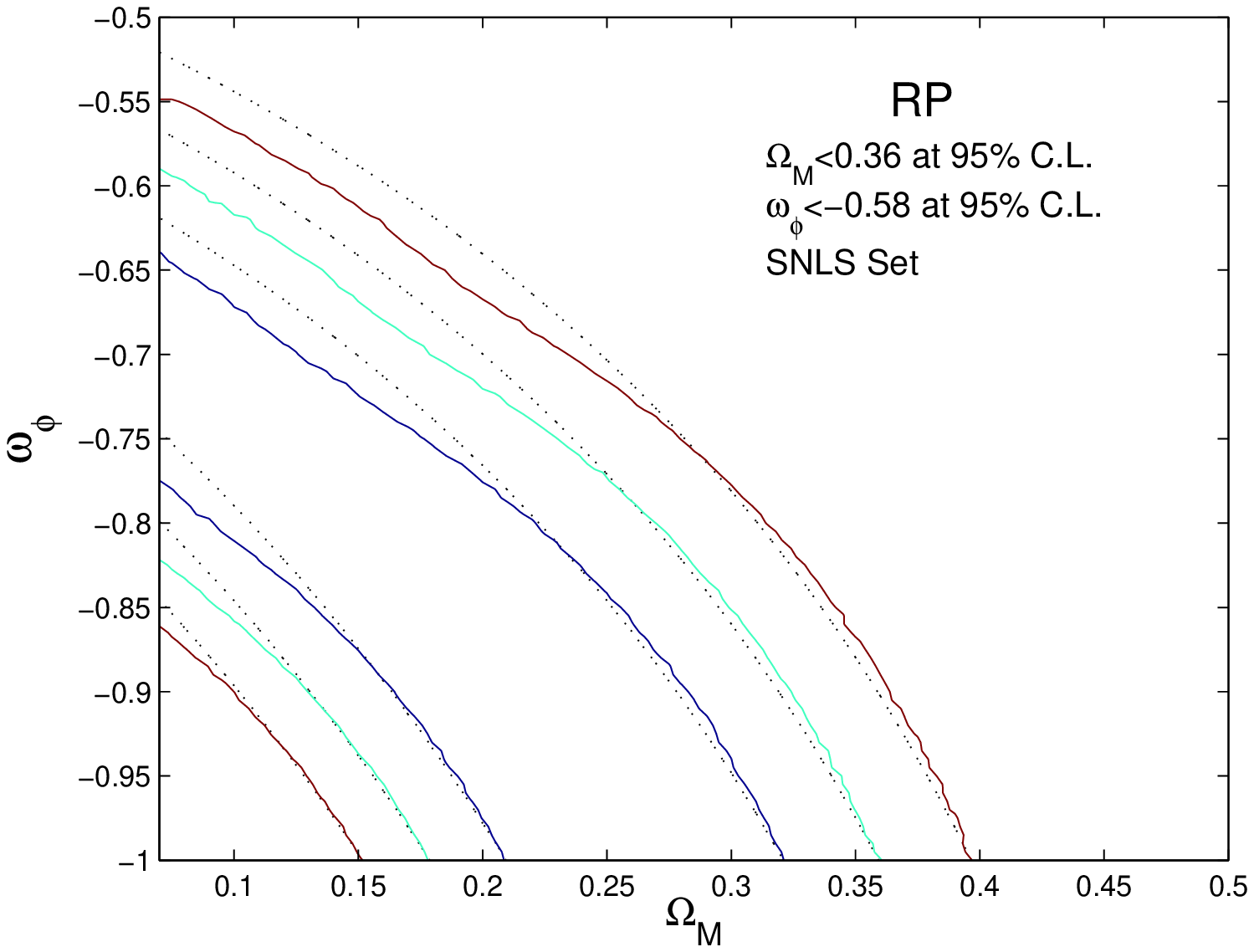}
\caption{ Confidence regions at the $68\%$, $95\%$ and $99\%$  C.L.
(corresponding to $\Delta\chi^{2}$=2.3, 6 and 11.8 for a two-parameter fit) 
for ($\alpha,\Omega_{M}$) (left panels) and ($\omega_{\phi},\Omega_{M}$) 
(right panels), in the Ratra-Peebles quintessence potential, 
obtained using the Gold (top panels) and the SNLS 
(bottom panels) samples. The dotted lines, on the right panels, 
show the probability contours for a XCDM model: at the $99\%$ 
and $95\%$ for the Gold set, 
and at the $99\%$, $95\%$, and $68\%$ for the SNLS set.}
\end{figure*}

\begin{figure*}[h]
\includegraphics[width=0.49\textwidth]{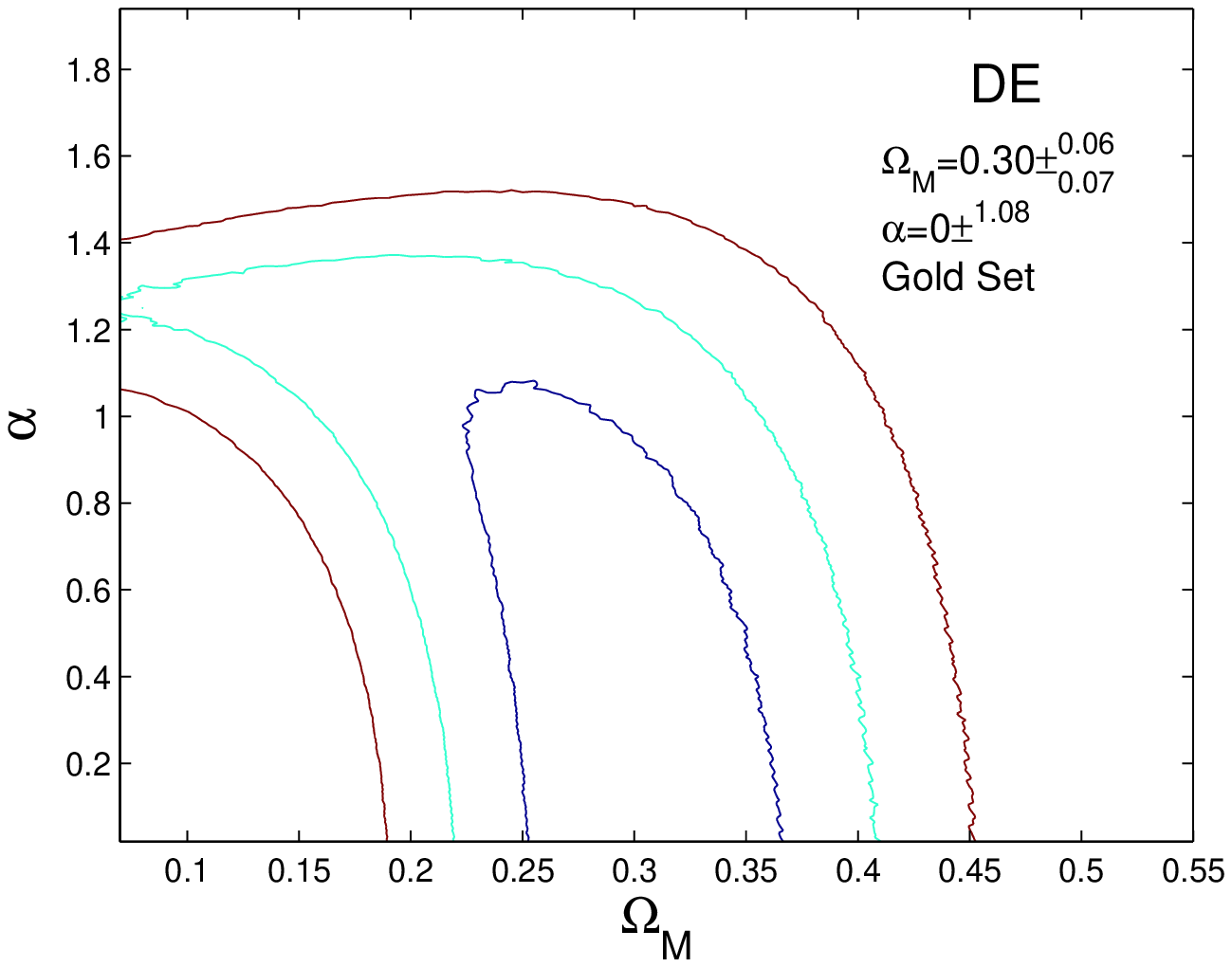}
\includegraphics[width=0.49\textwidth]{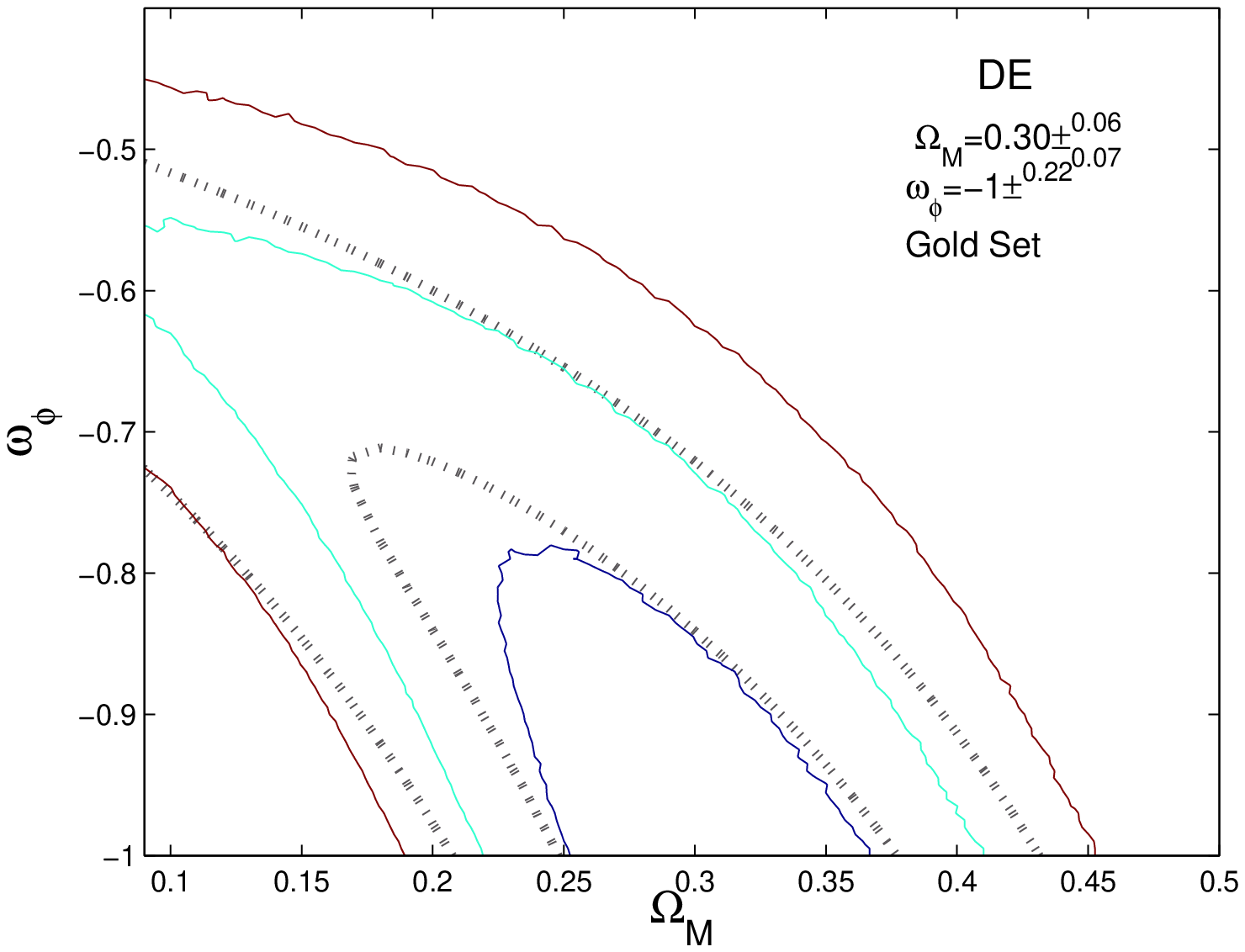}
\includegraphics[width=0.49\textwidth]{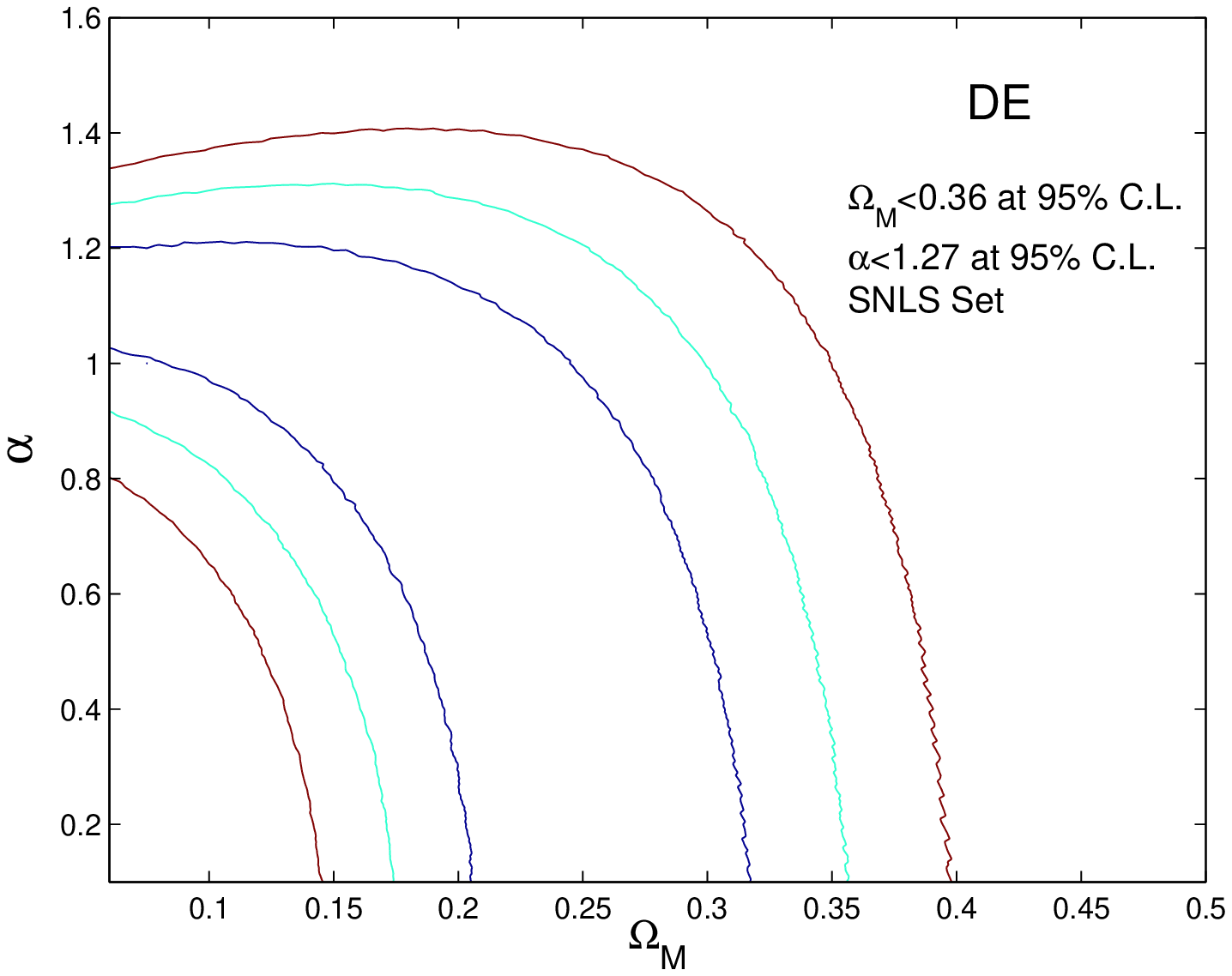}
\includegraphics[width=0.49\textwidth]{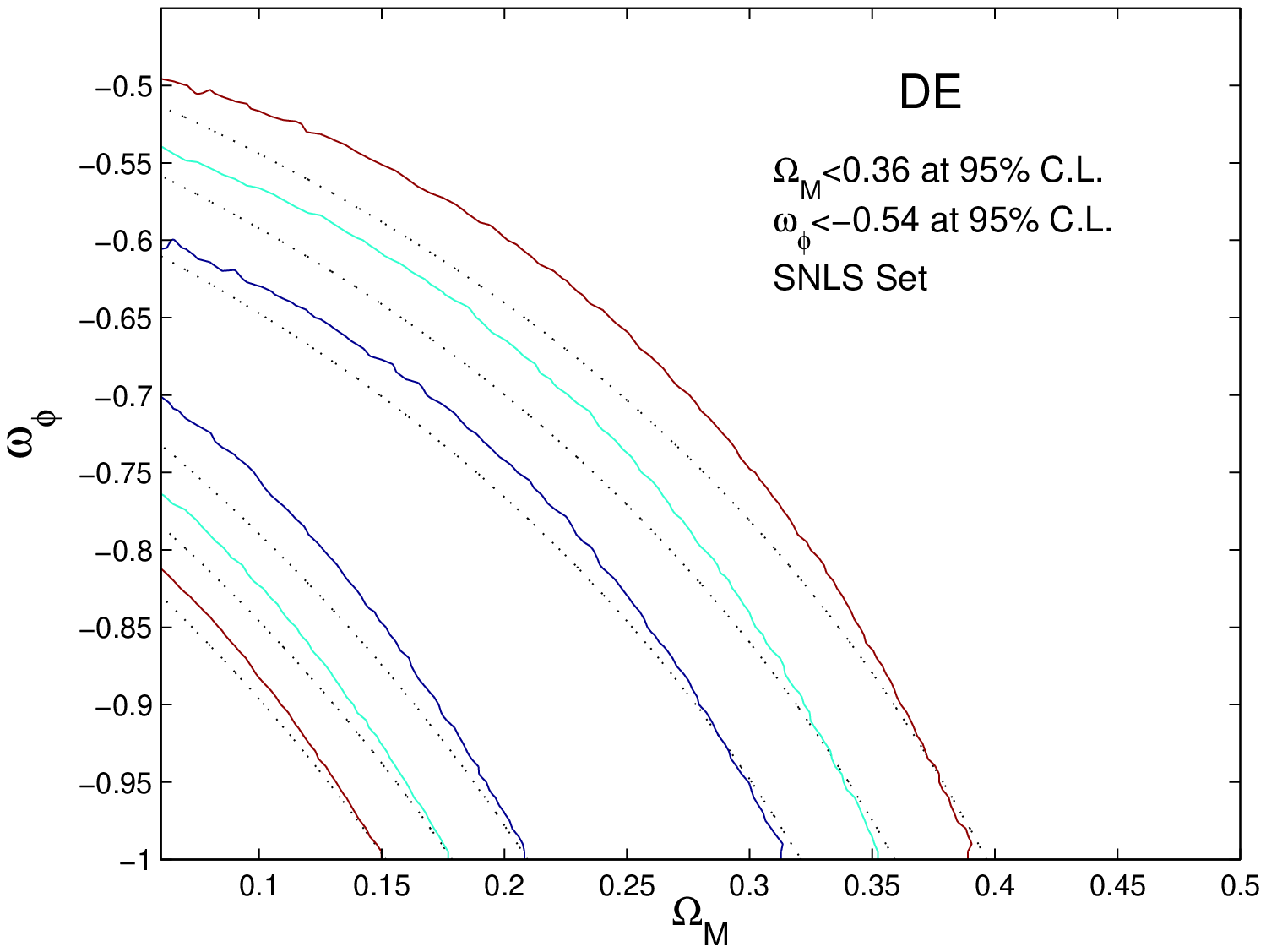}
\caption{ Confidence regions at the $68\%$, $95\%$ and $99\%$ C.L. 
(corresponding to $\Delta\chi^{2}$=2.3, 6 and 11.8 for a two-parameter fit) 
for ($\alpha,\Omega_{M}$) (left panels) and ($\omega_{\phi},\Omega_{M}$) 
(right panels), in the Double Exponential quintessence potential, obtained 
using the Gold (top panels) and the SNLS (bottom panels) samples. 
The dotted lines, on the right panels, show the probability contours 
for a XCDM model: at the $99\%$ and $95\%$ for the Gold set, 
and at the $99\%$, $95\%$, and $68\%$ for the SNLS set.}
\end{figure*}

\begin{figure*}[h]
\includegraphics[width=0.49\textwidth]{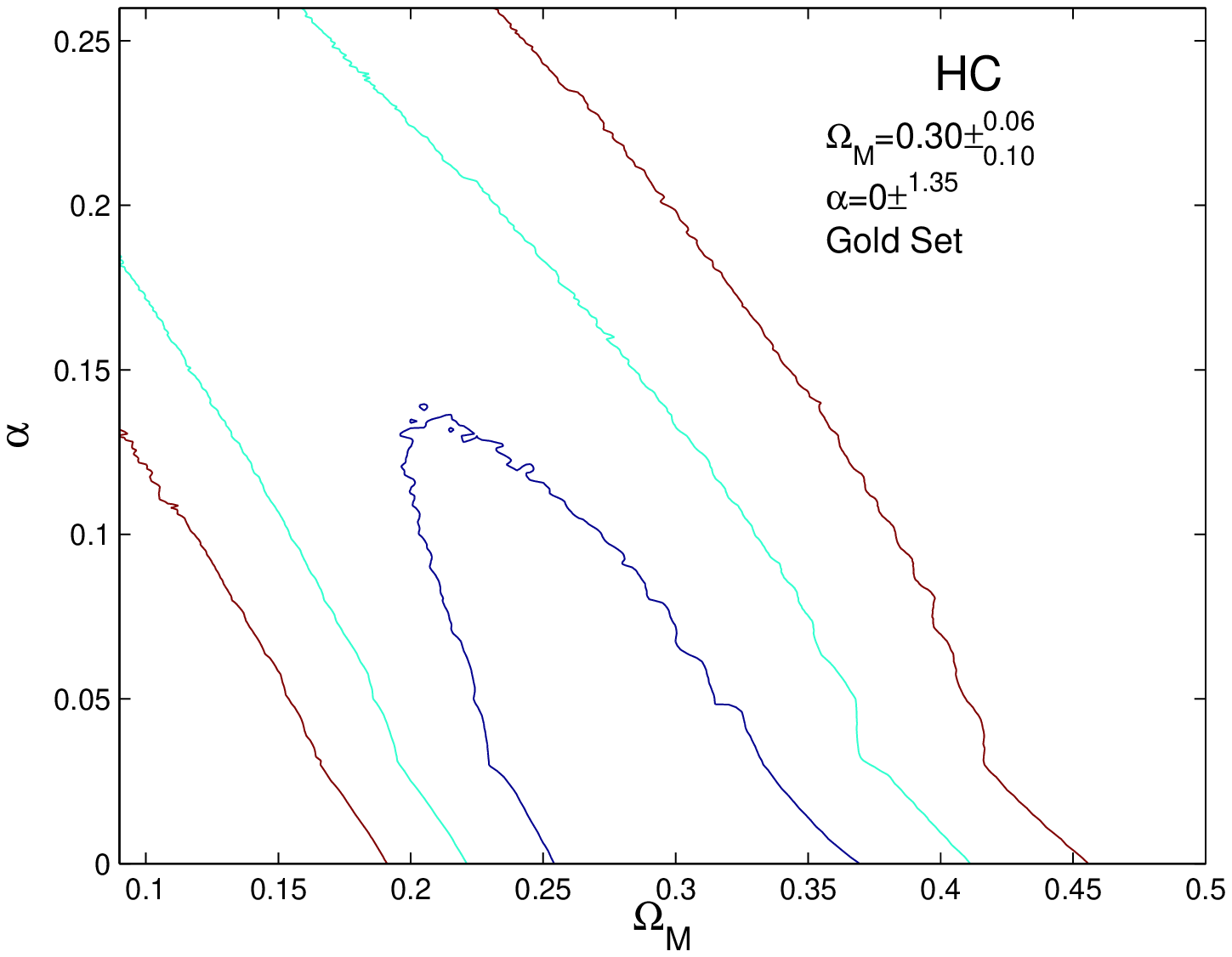}
\includegraphics[width=0.49\textwidth]{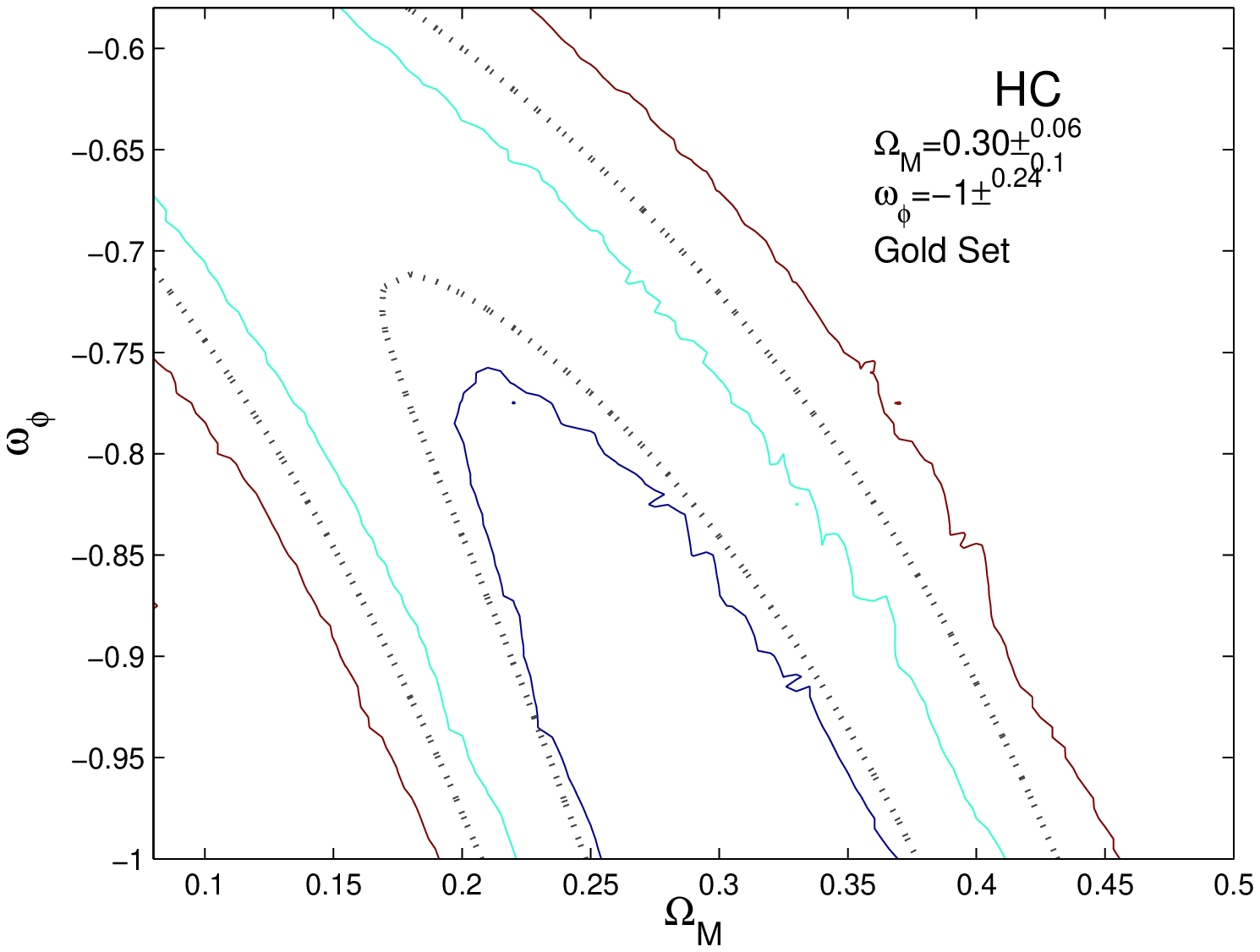}
\includegraphics[width=0.49\textwidth]{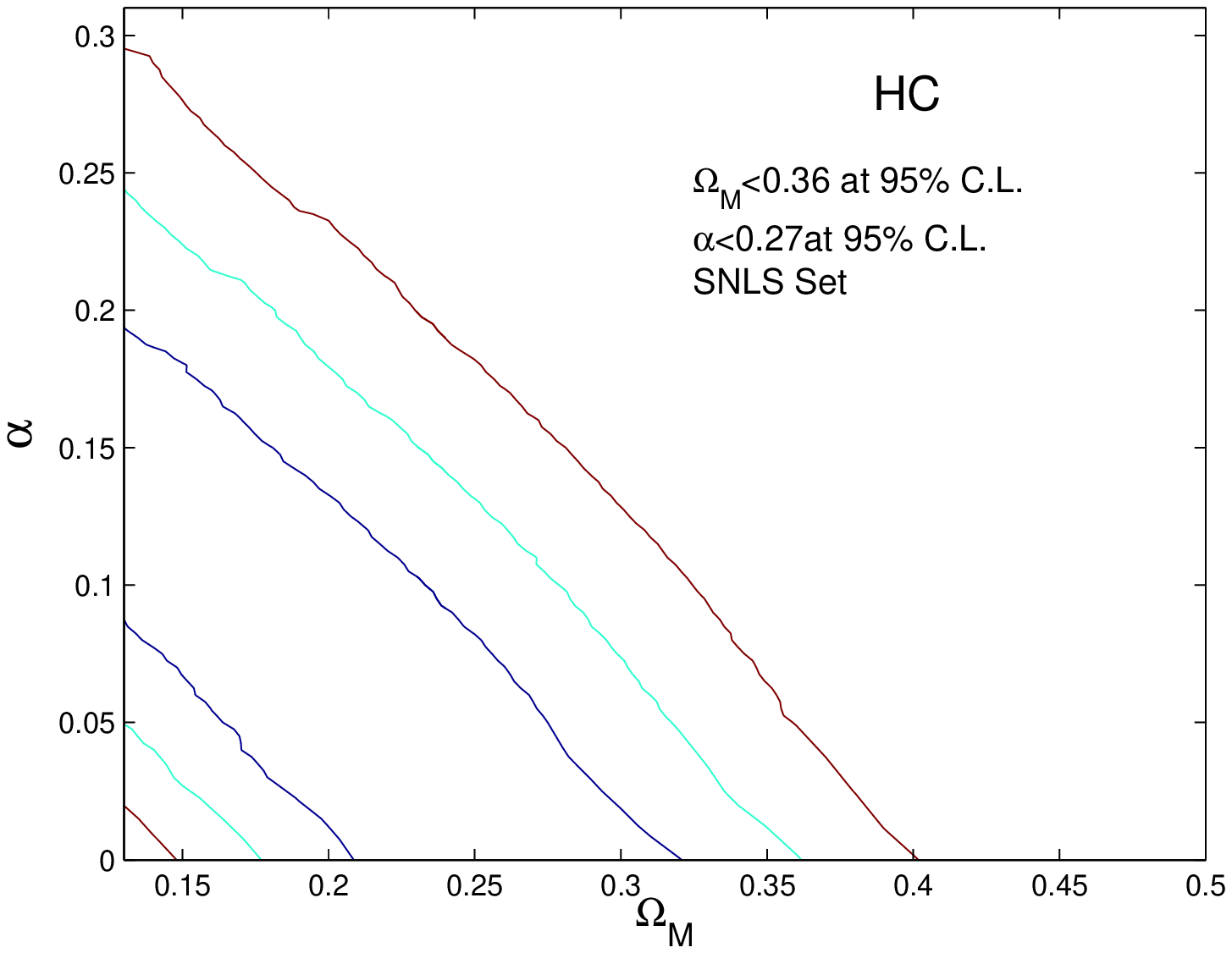}
\includegraphics[width=0.49\textwidth]{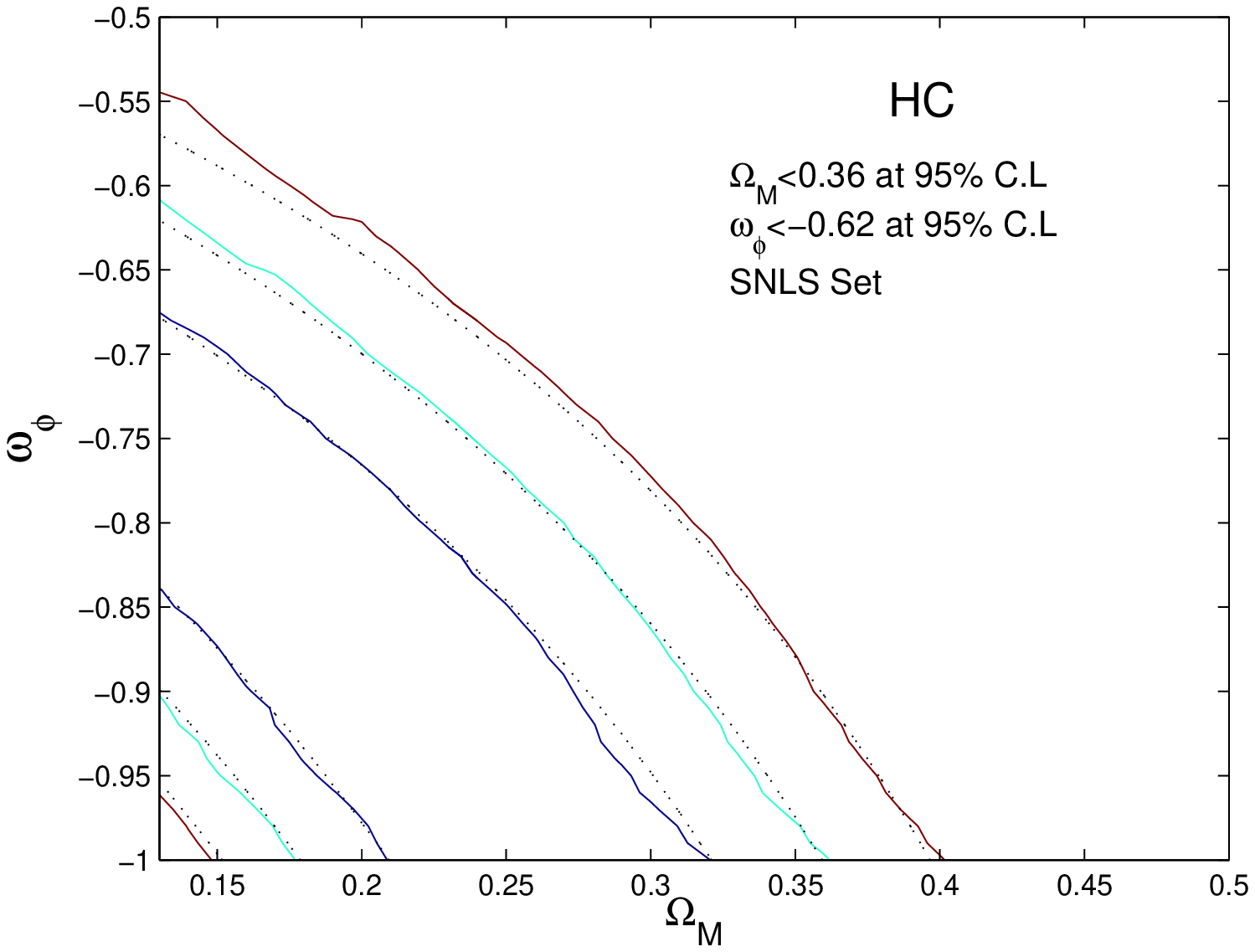}
\caption{  Confidence regions at the $68\%$, $95\%$ and $99\%$ C.L.   
(corresponding to $\Delta\chi^{2}$=2.3, 6 and 11.8 for a two-parameter fit) 
for ($\alpha,\Omega_{M}$) (left panels) ($\omega_{\phi},\Omega_{M}$) 
(right panels), in the Hyperbolic Cosine quintessence potential, 
obtained using the Gold (top panels) and the SNLS (bottom panels) 
samples. The dotted lines, on the right panels, show the probability 
contours for a XCDM model: at the $99\%$ and $95\%$ for the Gold 
set, and at the $99\%$, $95\%$, and $68\%$ for the SNLS set.}
\end{figure*}

\section{Conclusions}
We have compared the predictions of three quintessence models with the
measurements of the luminosity distance vs. redshift from two SNe Ia catalogues,
 (Gold and SNLS surveys). We have obtained the corresponding 
confidence regions in the  ($\alpha,\Omega_{M}$) and 
($\omega_{\phi},\Omega_{M}$) planes and compared them with the 
predictions of 
dark energy models with constant equation of state (XCDM). 
From the plots we see that
the different confidence regions for quintessence and XCDM models cover
essentially the same areas in the ($\Omega_M,w_\phi$) plots. This makes
it difficult to discriminate between the different models just 
from SNe Ia data alone.  
In all the
three cases, 
the best fit cosmology corresponds to the $\alpha=0$ case, in which 
the quintessence potential reduces effectively to a cosmological constant.
Although SNLS data favor a pure cosmological constant, they 
yield larger confidence regions than the Gold dataset. 
Bounds on the cosmological parameters $\Omega_{M}$, $\omega_{\phi}$ and
the potential paramaters $\alpha$ 
have also been obtained. In 
principle it could be interesting to use 
the differences in constraints between the two sets 
as an estimate of possible residual systematics in the 
SNe Ia data. However, in practice, since both datasets have been 
obtained and reduced in different ways and possibly the actual 
data is still dominated by statistical errors, it would be
difficult to perform such estimations. In any case this is beyond
the scope of the present work.

\vspace{.5cm}

{\em Acknowledgments:} 
We would like to thank P. Astier for useful comments. 
This work has been partially supported by DGICYT (Spain) under project numbers
FPA 2004-02602 and FPA 2005-02327


\begin{thebibliography}{999}
\bibitem{Gold} A.G. Riess at al. {\it Astrophys.J.} {\bf 607}, 665 
(2004) 
\bibitem{SNLS} P. Astier et al.,  astro-ph/0510447
\bibitem{WMAP} D.N. Spergel {\it et al.}, 
  {\it Astrophys. J. Suppl.} {\bf 148}, 175 (2003)
\bibitem{review}  P. J. E. Peebles and B. Ratra, {\it Rev. Mod. Phys.} {\bf 75}, 559 (2003)
\bibitem{review2} E. J. Copeland, M. Sami and S. Tsujikawa, hep-th/0603057
\bibitem{RP} B. Ratra and P.J.E. Peebles, {\it Phys. Rev.} {\bf D37},
3406 (1988)
\bibitem{quintessence} C. Wetterich, {\it Nucl. Phys.} {\bf B302}, 668 (1988); 
R.R. Caldwell, R. Dave and P.J. Steinhardt, {\it Phys. Rev. Lett.}
 {\bf 80}, 1582 (1998)
\bibitem{tracking} P.~J.~Steinhardt, L.~M.~Wang and I.~Zlatev,
 {\it  Phys. Rev.} {\bf D59} 123504 (1999) 
\bibitem{scaling} E.J Copeland, A.R. Liddle and D. Wands, {\it Phys. Rev.} {\bf D57}, 4686 (1998) 
\bibitem{scaling2} P. G. Ferreira and M. Joyce,  {\it Phys. Rev.} {\bf D58}, 023503 (1998) 
\bibitem{scaling3} S. Mizuno, S. J. Lee and E. J. Copeland, { \it Phys.
Rev.} {\bf D70}, 043525 (2004); E. J. Copeland, S. J. Lee,
J. E. Lidsey and S. Mizuno, {\it Phys. Rev.} {\bf D71}, 023526
(2005); M. Sami, N. Savchenko and A. Toporensky, { \it Phys.
Rev.} {\bf D70}, 123526 (2004); V. Pettorino, C. Baccigalupi
and F. Perrotta, JCAP {\bf 0512}, 003 (2005); F.~Piazza and S.~Tsujikawa,
JCAP {\bf 0407}, 004 (2004); S.~Tsujikawa and M.~Sami, 
{\it Phys. Lett.} {\bf B603}, 113 (2004)

\bibitem{k-essence} C. Armendariz-Picon, T. Damour and V. Mukhanov, {\it Phys. Lett.} {\bf B458}, 209 (1999); T. Chiba, T. Okabe and M. Yamaguchi, 
{\it Phys. Rev.} {\bf D62}, 023511 (2000)
\bibitem{Chaplygin} A.Y. Kamenshchick, U. Moschella and V. Pasquier, 
{\it Phys. Lett.} {\bf B511}, 265 (2001); N. Bilic, G.B. Tupper and 
R.D. Viollier, {\it Phys. Lett.} {\bf B535}, 17 (2002); M.C. Bento, 
O. Bertolami and A.A. Sen, {\it Phys. Rev.} {\bf D66}, 043507 (2002)
\bibitem{infrared} G. Dvali, G. Gabadadze and M. Porrati, 
{\it Phys. Lett.} {\bf B485}, 208 (2000); 
S. Capozziello, S. Carloni and A. Troisi, astro-ph/0303041.
\bibitem{Cardassian} K. Freese and M. Lewis, 
{\it Phys. Lett.} {\bf B540}, 1 (2002); S. Sen and A. A. Sen {\it Astrophys.J.} {\bf 588}, 1 
(2003) 
\bibitem{Bento} M.C. Bento, O. Bertolami, N.M.C. Santos and
 A.A. Sen, {\it Phys. Rev.} {\bf D71}, 063501 (2005)
 \bibitem{ness1}S. Nesseris, L. Perivolaropoulos {\it Phys.Rev.} 
{\bf D70} 043531 (2004)
\bibitem{pref_phantom}U. Alam, V. Sahni, T. D. Saini, and A. A. Starobinsky, {\it Mon. Not. R. Astron. Soc.} {\bf 354}, 275 (2004); 
T. R. Choudhury and T. Padmanabhan, {\it Astron. Astrophys.} {\bf 429}, 807 (2005);
Y. Wang and P. Mukherjee, {\it Astrophys. J.} {\bf 606}, 654 (2004); 
D. Huterer and A. Cooray, {\it Phys. Rev.} {\bf 71}, 023506 (2005); 
\bibitem{Nesseris} S.Nesseris, L. Perivolaropoulos, {\it Phys. Rev.} {\bf D72}, 123519 (2005),
\bibitem{pref_LCDM} H. K. Jassal, J. S. Bagla, T. Padmanabhan,
 astro-ph/0601389
\bibitem{alter1} R.R. Caldwell and M. Doran, {\it Phys. Rev.} 
{\bf D69}, 103517 (2004)
\bibitem{alter} M. Doran, M.J. Lilley, J. Schwindt and C. Wetterich,
 {\it Astrophys. J.} {\bf 559}, 501 (2001); 
M.S. Movahed and S. Rahvar, {\it Phys. Rev.} {\bf D73}, 083518 (2006) 

\bibitem{DEperturbations} W. Hu. {\it Astrophys. J.} {\bf 506}, 485 (1998);
T. Moroi and T. Takahashi, {\it Phys. Rev. Lett.} 
{\bf 92}, 091301, (2004); C. Gordon and W. Hu, {\it Phys. Rev.}
{\bf D70}, 083003, (2004); A.L. Maroto, {\it JCAP} {\bf 0605}:015 (2006)  
\bibitem{DE} T. Barreiro, E. J. Copeland and N.J. Nunes,
 {\it Phys. Rev.} {\bf D61}, 127301 (2000)
\bibitem{DE2} A. A. Sen and S. Sheti , {\it Phys.Rev.} {\bf B532}, 159 (2002);
I.P. Neupane, {\it Class. Quant. Grav.} {\bf 21}, 4383 (2004);
I.P. Neupane, {\it Mod. Phys. Lett.} {\bf A19}, 1093 (2004);
L. Jarv, T. Mohaupt and F. Saueressig, JCAP {\bf 0408}, 016 (2004)
\bibitem{HC}  V. Sahni and L. M. Wang {\it Phys. Rev.} {\bf D62}, 103517 
(2000) 
\bibitem{param_w} R. Lazkoz, S. Nesseris, L. Perivolaropoulos, JCAP {\bf 0511}, 010 (2005);
 V. Barger, E. Guarnaccia, D Marfatia, {\it Phys. Lett.} {\bf B635}, 
61-65 (2006)
\bibitem{ichikawa} K. Ichikawa and T. Takahashi,
{\it Phys. Rev.} {\bf D73}, 083526 (2006); Z.K. Guo, N. Ohta and
Y.Z. Zhang, {\it Phys. Rev.} {\bf D72}, 023504 (2005)  
\bibitem{bassett} B.A. Bassett, P.S. Corasaniti and M. Kunz,
{\it Astrophys. J.} {\bf 617}, L1 (2004)
\bibitem{caresia} P. Caresia, S.Matarrese and L.Moscardini,  
{\it Astrophys. J.} {\bf 605}, 21-29 (2004)


\end{thebibliography}
\end{document}